\documentclass{article}
\usepackage[linesnumbered,ruled]{algorithm2e}
\usepackage{graphicx}
\usepackage{url}
\pdfoutput=1

\title{
  Simulating fluids with a computer: \\ 
  Introduction and recent advances
}
\author{Bruno L\'evy}

\begin{document}

\maketitle

\begin{abstract}

In this article, I present recent methods for the numerical simulation
of fluid dynamics and the associated computational algorithms. The
goal of this article is to explain how to model an incompressible
fluid, and how to write a computer program that simulates it.  I will
start from Newton laws ``${\bf F} = m {\bf a}$'' applied to a bunch of
particles, then show how Euler's equation can be deduced from them by
``taking a step backward'' and seeing the fluid as a continuum. Then I
will show how to make a computer program. Incompressibility is one of
the main difficulties to write a computer program that simulates a
fluid. I will explain how recent advances in computational mathematics
result in a computer object that can be used to represent a fluid and
that naturally satisfies the incompressibility constraint. Equipped
with this representation, the algorithm that simulates the fluid
becomes extremely simple, and has been proved to converge to the
solution of the equation (by Gallouet and Merigot).  

\end{abstract}

\begin{figure}
  \centerline{
    \includegraphics[width=0.22\textwidth]{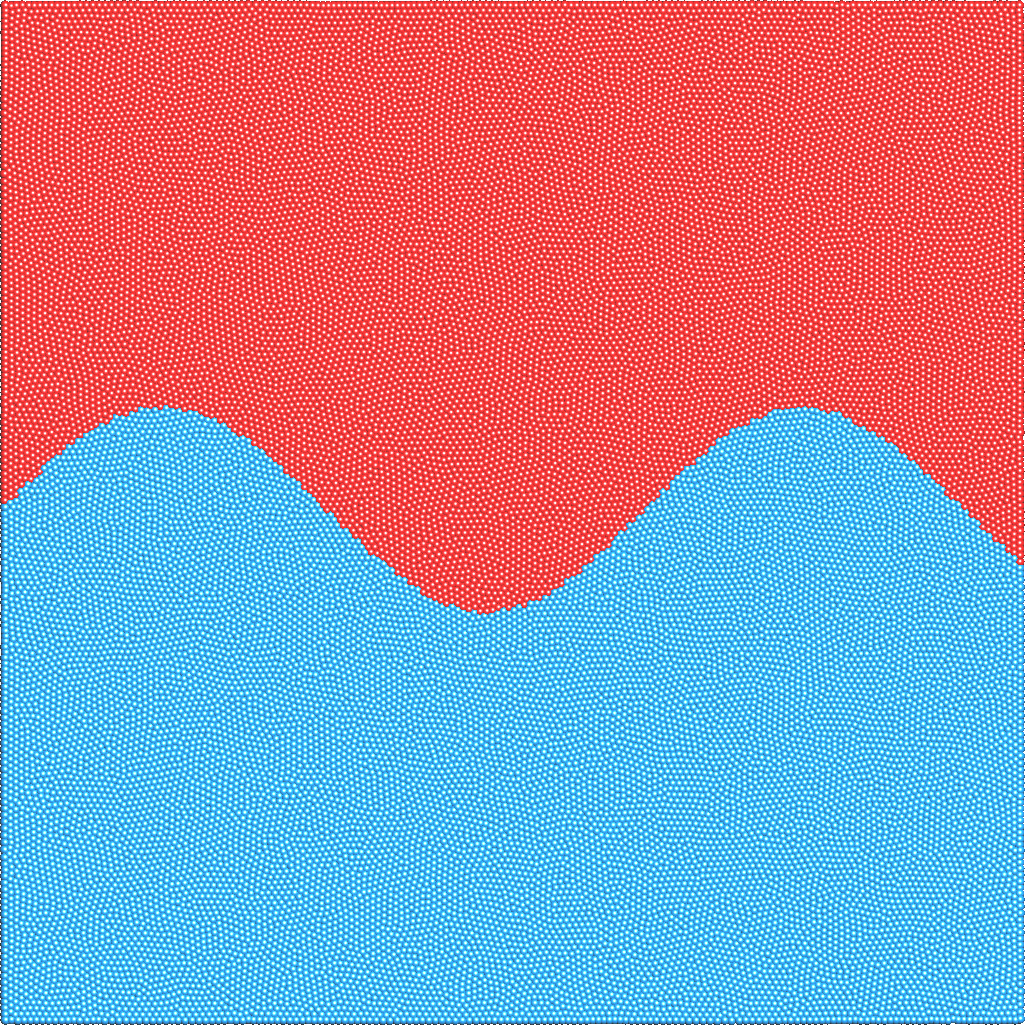}
    \includegraphics[width=0.22\textwidth]{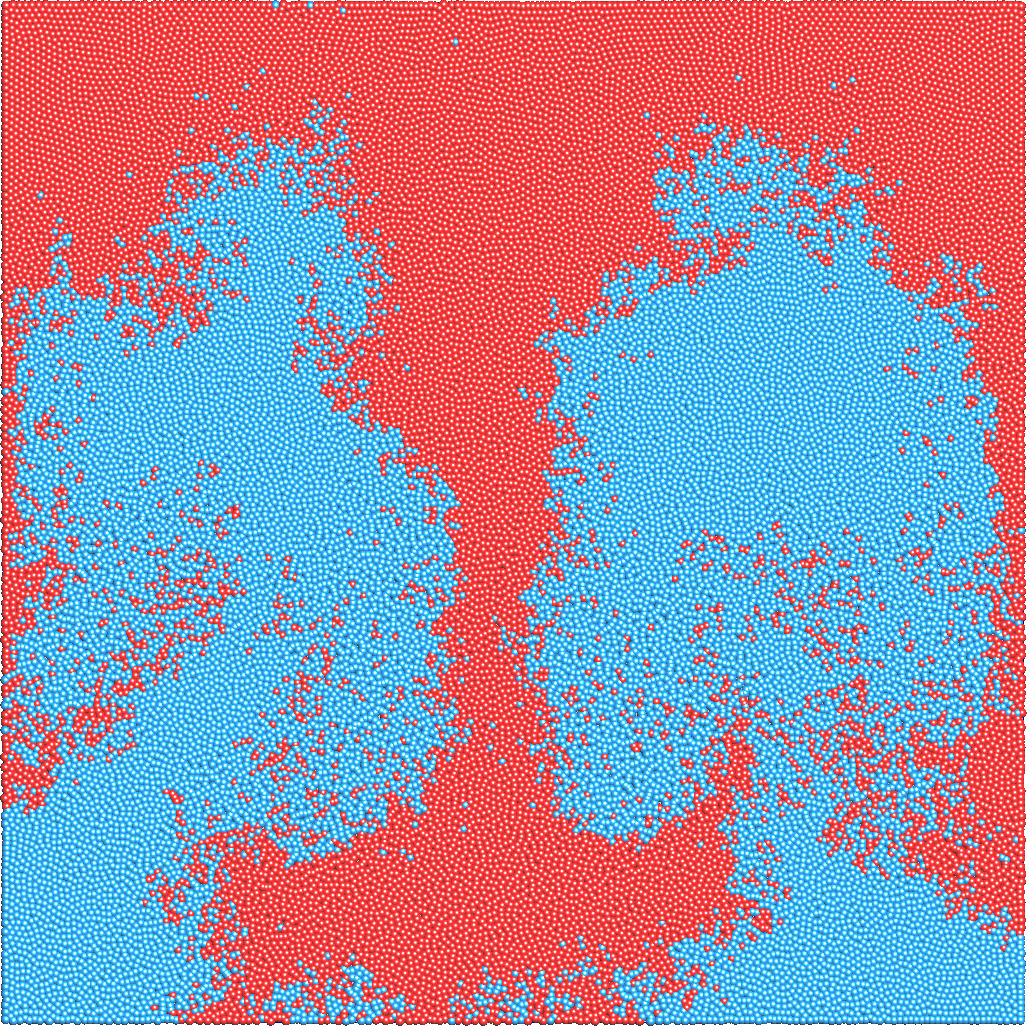}
    \includegraphics[width=0.22\textwidth]{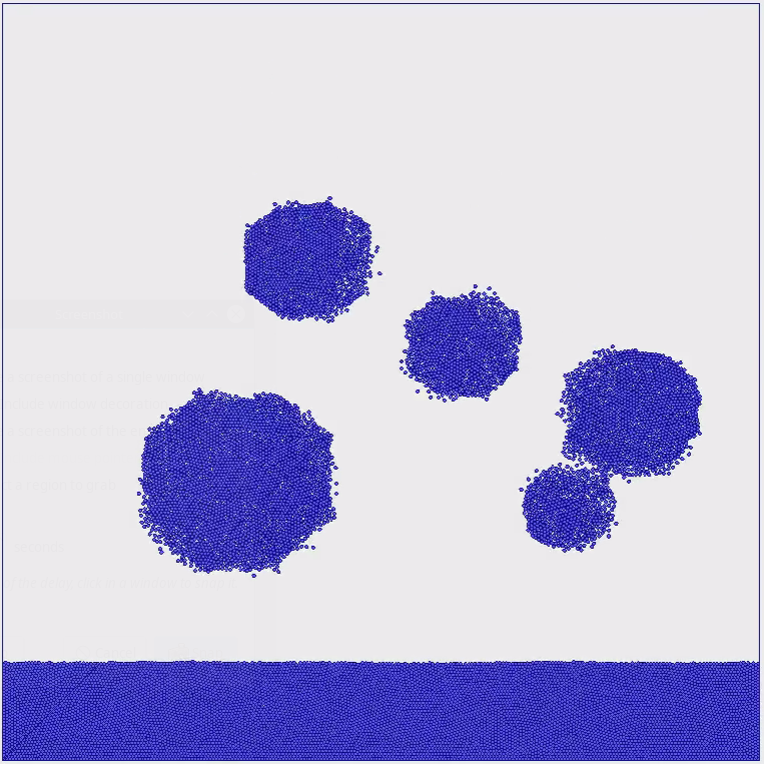}
    \includegraphics[width=0.22\textwidth]{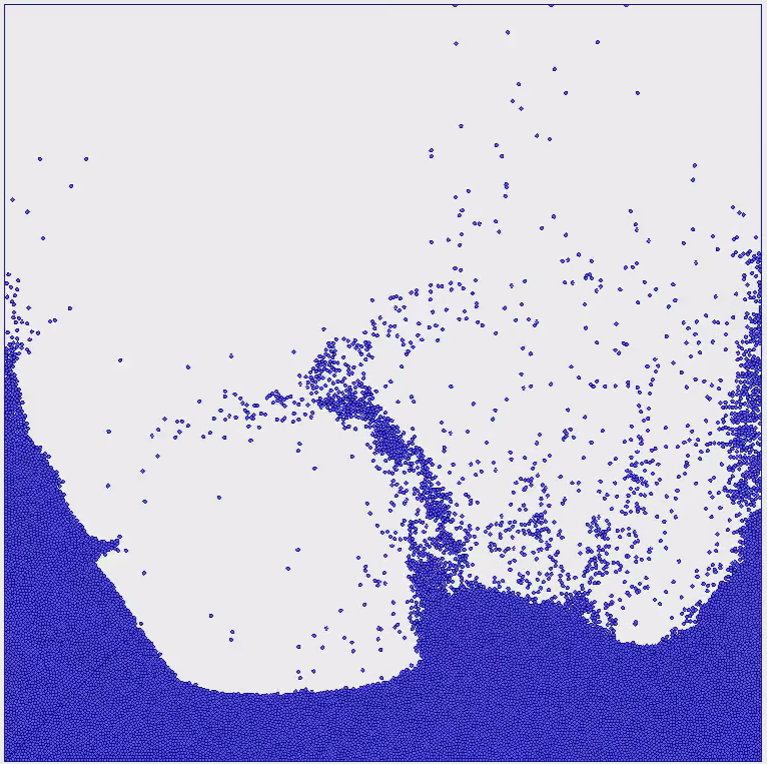}
  } 
  \caption{Interesting behaviors of fluids (simulated). Left: starting from a configuration with
    a heavy fluid (water, in red) on top of a light fluid (oil, in blue), both fluids
    want to exchange their positions. Because fluids are incompressible, there is no trivial
    path to exchange their positions, thus some nice vortices appear. Right: free-surface fluid
    simulations with some drops of water. This creates changes of shape and topology, as well as splashes.}
  \label{fig:fluids}
\end{figure}

\section*{Introduction}
Fluids are objects that are notoriously difficult to simulate on a computer. I think
the difficulty comes from different aspects, concerning both the nature of the object,
the physics, and the mathematical modeling:
\begin{enumerate}
\item {\bf Shape and topology of interfaces:} during its movement, a
  fluid can show drastic changes of shape and topology.  The free
  surface of a fluid can split and merge, create splashes, \ldots When
  considering two non-miscible fluids (water and oil), the interface
  between both fluids can form complicated shapes (see Figure \ref{fig:fluids});
\item {\bf conservation of physical quantities:} it is well known that
  some quantities are conserved, the most obvious one being mass. The
  kinetic energy ($1/2 m {\bf v}^2$) is also conserved, as well as
  momentum ($ m {\bf v}$), and another vector quantity
  that corresponds to the way things are spinning (called ``angular momentum'').
  If considering an incompressible fluid (like most liquids), then the volume of a fluid
  element needs to be conserved as well. These conservation laws are
  not trivial to enforce in a numerical simulation;
\item {\bf form of the equations:} last but not least, the equations of fluids
  (Euler, Navier-Stokes) are written in a form that makes it difficult to
  understand their connection with the physics (Newton laws).
\end{enumerate}  

In what follows, I will first try to ``decipher'' the fluid equations (you will see them later)
by starting from the beginning, i.e., explaining how to construct them
from Newton laws (\S\ref{sec:FluidMath}). In the
same section, I will explain how this relates with problems
in astrophysics. Then I will briefly explain the existing methods
to implement a computer program that simulates a fluid using these
equations (\S\ref{sec:FluidNum}), as well as new ones, that use
recent advances in computational mathematics to deal with the incompressibility
constraint. Finally I will show some results (\S\ref{sec:FluidResults}).

\section{Mathematical modeling of fluids}
\label{sec:FluidMath}

\subsection{Back to the roots: Newton laws}

Since a fluid can be considered as a (huge !) number of particles in interaction, let us start with
the motion of a single particle, governed by Newton laws. Newton laws can be summarized as follows:
\begin{itemize}
\item {\bf Law (I) Inertia:} if there is no force, a particle continues its movement in a straight
  line, with a constant speed;
\item {\bf Law (II) Effect of forces:} if there is a force, then it changes the speed of the particle (accelerates/decelerates).
  The change of speed is proportional to the force, and inversely proportional to the mass of the particle. In formula:
  $$
     {\bf a} = \frac{1}{m}{\bf F}
  $$
     where $m$ denotes the mass of the particle, ${\bf F}$ the force and ${\bf a}$ the acceleration of the particle\footnote{It was initially
     written differently by Newton, without any reference to mass, he only stated that acceleration was proportional to force.}.
     Considering for instance gravity, the force (weight) is equal to the mass times gravity,
  ${\bf F} = m {\bf G} = m \begin{tiny} \left[ \begin{array}{c} 0 \\ 0 \\ -9.81 \end{array} \right] \end{tiny}$. Computing the acceleration
     ${\bf a}$, the mass cancels out, then you can deduce that if you fall, each second your speed gains 9.81 m/s, whatever your mass. \\
     Some forces, like gravity, can be completely deduced from a scalar field $\Phi$ (called ``potential'') and correspond to (minus) the
     gradient $\nabla \Phi = \begin{tiny}\left[\begin{array}{c} \partial \Phi / \partial x \\ \partial \Phi / \partial y \\ \partial \Phi / \partial z \end{array}\right]\end{tiny}$.
     For gravity on earth (considered to be flat !), with the Z axis in the vertical direction, $\Phi(x,y,z) = 9.81 m z$. At this point, it is interesting
     to notice that Newton's law connects time derivatives (acceleration) with space derivatives (gradient);
\item {\bf Law (III) Action and reaction:} consider two particles A and B. If A exerts a force ${\bf F}$ on B, then B exerts the
  force ${\bf -F}$ on A (same magnitude, opposite direction)\footnote{If you consider gravity, the force that attracts you to the earth,
    you exert to the earth exactly the same
  force (with a 'minus' sign). Then, why doesn't the earth move up when you jump ? This is because law (II) state that the ``change of speed''
  (acceleration) of the earth will be inversely proportional to the mass of the earth. In fact when you jump, the earth moves up a tiny bit, proportionally to the
  ratio of your mass and the mass of the earth.}. From this law, Newton deduced that for a set of particles in interaction, the vector
  ${\bf p} = \sum m_i {\bf v_i}$ remains constant, where $m_i$ and ${\bf v_i}$ denotes the mass and speed vector of each particle. The
  vector ${\bf p}$ is called ``momentum''\footnote{for French readers, ``quantit\'e de mouvement'' which means in English ``quantity of movement'',
    as in Newton's ``Principia Mathematica''. This term means that two particles of one gram moving at 1 meter per second in the same direction
    bear as much as ``quantity of movement''
    as one particle of two grams moving at 1 meter per second in that direction, or one particle of two grams moving at 2 meters per second.}.
\end{itemize}  
\emph{Side note: In case you expect it to come here, kinetic energy} $\frac{1}{2} m {\bf v}^2$
\emph{and conservation of energy are more subtle notions, difficult to connect to Newton's law.
  They were introduced by Leibniz who was contemporary of Newton. It took time for these notions
  to make their way, mainly due to the personality of Newton ! More on this here\footnote{\tiny \url{https://physics.stackexchange.com/questions/132754/how-was-the-formula-for-kinetic-energy-found-and-who-found-it}}}

\begin{algorithm}[t]
\KwIn{Initial positions ${\bf x}_i$ and speeds ${\bf v}_i$ of the $N$ particles}  
  $t \leftarrow 0$\;
  \For{{\rm timestep}=0 {\bf to} {\rm maxstep}} { 
    $t \leftarrow t + \delta t$ \;
    \For{i=1 {\bf to} N} {                       
       ${\bf x}_i \leftarrow {\bf x}_i + \delta t\ {\bf v}_i$  \tcp*{Update position}
    }
  }
\caption{Trivial simulation algorithm, no forces}
\label{alg:newton0}
\end{algorithm}

Suppose now that you want to simulate a bunch of $N$ particles on a computer. Each particle will be represented by a set of variables, first the
position ${\bf x}_i$ of the particle. Suppose (for now) that there is no force. Because of Law (I) (inertia), we need to memorize the speed 
${\bf v}_i$ of each particle. See the resulting (uninteresting) algorithm \ref{alg:newton0}.

\begin{algorithm}[t]
\KwIn{Initial positions ${\bf x}_i$ and speeds ${\bf v}_i$ of the $N$ particles}  
  $t \leftarrow 0$\;
  \For{{\rm timestep}=0 {\bf to} {\rm maxstep}} { 
    $t \leftarrow t + \delta t$ \;
    \For{i=1 {\bf to} N} {
       ${\bf F}_i \leftarrow m_i {\bf G}                   $  \tcp*{Update force}
    }
    \For{i=1 {\bf to} N} {    
       ${\bf a}_i \leftarrow \frac{1}{m_i}{\bf F}_i        $  \tcp*{Update acceleration}
       ${\bf v}_i \leftarrow {\bf v}_i + \delta t\ {\bf a}_i$  \tcp*{Update speed}      
       ${\bf x}_i \leftarrow {\bf x}_i + \delta t\ {\bf v}_i$  \tcp*{Update position}
    }
  }
\caption{Simple simulation algorithm with gravity}
\label{alg:newton1}
\end{algorithm}

Algorithm \ref{alg:newton0} is not very interesting, because all particles move in straight lines
at constant speed. To make it more interesting, we introduce forces. Let us consider for instance
gravity, in formula ${\bf F}_i = m{\bf G}$ where ${\bf G} = \begin{tiny} \left[ \begin{array}{c} 0 \\ 0 \\ -9.81 \end{array} \right] \end{tiny}$. At each
timestep, Algorithm \ref{alg:newton1} computes the force ${\bf F}_i$ applied to each particle, then
computes the acceleration (Newton law II), then updates the speed by integrating the acceleration,
and in turn updates the position by integrating the speed\footnote{There exist more elaborate ways
  of integrating w.r.t. time, such as Runge-Kutta, that uses higher-order polynomials during a timestep \cite{RK}, or
  Verlet, defined in such a way that the kinetic energy is conserved \cite{PhysRev.159.98}. In the frame of this
  article, to make things easier, I will continue using the simplest scheme, that multiplies the integrated quantity by $\delta t$
  (the so-called ``explicit Euler'' time integration).
}. Forces are updated in a different loop, because when considering more interesting forces, that involve particles interaction,
you may need the positions of all the particles at the previous timestep. If you implement this algorithm with a set of random
positions and random speed vectors, you will see that each particle moves along a parabola, as expected. 

Algorithm \ref{alg:newton1} is naive, but it is quite easy to extend
it to other types of forces. In fact, this is how Newton discovered
his laws, trying to figure out what happens to a planet that orbits around the sun
when moving from $t$ to $t + \delta t$, and how ${\bf x}$ moves from ${\bf x}$ to ${\bf x} +
\delta {\bf x}$, in a way that some known laws of motion (Kepler laws) could be retrieved.
Newton's and Leibniz's genus was to consider what
happens when $\delta t$ tends to zero, and this is how they invented
the concept of derivative / differential calculus\footnote{Their contemporaries did not believe in these computations and
mocked them as ``ghosts of vanished quantities''.}. It is fortunate
that they lived in the 1700's. If they had a computer, they would
probably simply have used a tiny timestep to do the computations
numerically, without inventing the mathematical theory. In our case,
the theory gives us directly the equation of the parabola for each
particle, without needing to do this ``numerical
integration''. However, sometimes it is impossible to derive the
solution formally. For instance, in the early NASA space program,
solving the system of PDEs that governs the re-entry of the spaceship
required to use numerical integration. The ``Hidden Figures'' movie portrays
Katherine Goble Johnson, one of the ``human computers'' of the NASA, who played
a key role in this story\footnote{
  \url{https://www.insidescience.org/news/exploring-math-hidden-figures}
}.

\subsection{From Newton laws to incompressible Euler fluids}

\subsubsection*{From discrete to continuum}
We now consider a fluid, that is to say a huge number $N$ of particles. At each time $t$,
each particle has a position ${\bf x}_i(t)$, a speed ${\bf v}_i(t) = \frac{d {\bf x}_i}{d t}$
and an acceleration ${\bf a}_i(t) = \frac{d^2 {\bf x}_i}{d t^2}$. Newton's law (II) states that
${\bf a}_i(t) = \frac{1}{m_i}{\bf F}_i(t)$. Up to now, to name a particle, we use its index $i$,
which is good when we got a small number of particles, but remember that we got a huge number of
particles, so another possibility for the ``name'' of a particle is to use its position at time $t=0$.
Not only this makes it easier to know which particle you are talking about, but also this makes it
possible to talk about an infinite number of particles !

One way of doing that is to represent the set of all particle trajectories with a function 
${\bf \chi}({\bf x}_0, t) : \Omega \times [0,T] \rightarrow \Omega$ where $\Omega$ denotes
the geometric domain and $T$ the simulation time. If you know it, the function
${\bf \chi}({\bf x}_0, t)$ tells you where the point that was at position ${\bf x}_0$ at
time $0$ is at time $t$. Supposing that all the particles have a mass $m$, Newton's law (II) rewrites as:

\begin{equation}
  \forall {\bf x}_0 \in \Omega, \forall t \in [0,T], \quad \quad \frac{d^2 {\bf \chi}({\bf x}_0,t)}{d t^2} = \frac{1}{m} {\bf F({\bf \chi}({\bf x}_0,t)},t)
  \label{eqn:LagrangeNewton}
\end{equation}

  We also need to take into account the fact that fluids are incompressible. In other words, this means that if we consider an element of fluid
$B \subset \Omega$ at time $t=0$, then at any time it should have the same volume. Since this condition needs to be satisfied for any subset $B$
of $\Omega$, it needs to be satisfied for an elementary (arbitrarily small) volume around each point ${\bf x}_0$ of $\Omega$. The way ${\bf \chi}$
changes an elementary volume corresponds to the determinant of its Jacobian matrix (and in our case, it needs to remain constant):
$$
   \forall {\bf x}_0 \in \Omega, \forall t \in [0,T], \quad \quad \mbox{det}( J {\bf \chi} ) = \mbox{constant}
$$
where the Jacobian matrix $J {\bf \chi}$, taken with respect to space, is given by:
$$
J {\bf \chi} = \left[
  \begin{array}{ccc}
    \frac{\partial \bf \chi_x}{\partial x} & \frac{\partial \bf \chi_x}{\partial z} & \frac{\partial \bf \chi_x}{\partial y} \\
    \frac{\partial \bf \chi_y}{\partial x} & \frac{\partial \bf \chi_y}{\partial z} & \frac{\partial \bf \chi_y}{\partial y} \\
    \frac{\partial \bf \chi_z}{\partial x} & \frac{\partial \bf \chi_z}{\partial z} & \frac{\partial \bf \chi_z}{\partial y}
  \end{array}    
\right]
$$

There are several difficulties:
\begin{itemize}
\item first, if you imagine you are looking at the fluid, say a river, from above, sitting on a bridge that crosses the river, it
  will be difficult to keep track of this infinite number of particles that move along many different trajectories. It may be easier
  to measure the fluid through a fixed grid (you can think about it as a metallic net sitting on the banks of the river), and then
  measuring quantities such as the number of particles in each cell of the grid, and the speed vector of the particle that
  passes under a grid intersection;
\item second, to be compatible with the rest, the constraint on volume preservation needs to be expressed as an additional force,
  injected into the right hand side of Equation \ref{eqn:LagrangeNewton}. While it is possible to do that in the form of Equation
  \ref{eqn:LagrangeNewton}, it is easier to express in function of the grid mentioned above.
\end{itemize}  
  
\subsubsection*{Particle coordinates and field coordinates}

Equation \ref{eqn:LagrangeNewton} considers Newton's laws from the ``point of view'' of a particle (also called ``Lagrange'' point of view, or ``Lagrange coordinates'').
The ``name'' of the particle -- that is, the way it is referenced in the equation -- is ${\bf x}_0$, i.e., where the particle was at time $t=0$, and the function
${\bf \chi}({\bf x}_0, t)$ tells you where it is after time $t$.

Consider now the metallic grid attached to the banks of the river. Then there are several quantities of interest:
\begin{itemize}
\item at time $t$, you may count the particles that are in a given cell of the grid. Make the number of particles tend to infinity and the size of the grid
  cells tend to zero, then this number of particles becomes a density, that we will denote by $\rho({\bf x},t)$. Note: clearly, if the fluid is incompressible and density
  was uniform (i.e., constant w.r.t. space) at time $t=0$, then density remains constant and uniform;
\item still at time $t$, imagine that you are staring at a ``grid
  point'' ${\bf x}$ at the intersection of two wires of the fixed
  ``metallic grid'' attached to the banks of the river. We will denote
  by ${\bf u}({\bf x},t)$ the speed of the particle that passes
  exactly under grid point ${\bf x}$ at time t. Another way of thinking
  about ${\bf u}$ is imagining an array of ``weathercocks'' that measure
  both the direction and strength of the wind at a set of fixed locations.
\end{itemize}

By considering that the number of particles tends to infinity
and the size of the grid cells tends to zero, you obtain physical quantities attached to fixed positions in space,
and defining what is called a ``field'', such as the ``density field'' $\rho$, and the ``velocity field'' ${\bf u}$. The point
of view of fields, or of a person sitting on the bridge and looking at the fluid through a fixed grid,
is called ``Euler'' point of view, or ``Euler coordinates''.

Now the question becomes: ``how can we write Newton's equation in terms of $\rho$ and ${\bf u}$ ?'' (then it will become
``how can we express incompressibility as an additional force ?''. Before answering this question, some care needs to be taken:
is it possible for $\rho$ and ${\bf u}$ to take arbitrary values ? The answer is no: they are connected. Imagine you know
$\rho(.,t)$ and ${\bf u}(.,t)$ at a given time $t$. When you go from $t$ to $t + \delta t$, the new $\rho(.,t+\delta t)$
is the result of ``transporting''the matter with ${\bf u}(.,t)$, thus it cannot be arbitrary. Put differently, if you consider
a small blob around a point ${\bf x}$ at time $t$, the quantity of matter that enters the blob minus the quantity of matter that
leaves the blob (deduced from the spatial variations of ${\bf u}$) should correspond to the variation of the quantity of
matter inside the blob (the time variation of $\rho$). Write this condition, use the
divergence theorem, and make the blob arbitrarily small, then you end up with this condition:
$$
   \frac{\partial \rho}{\partial t} = -\mbox{div}({\rho {\bf u}}) = -\nabla \cdot (\rho {\bf u})
$$
where $\mbox{div} = \nabla\cdot$ denotes the divergence. This condition corresponds to mass conservation (and is called the
``continuity'' equation, referring to the fact that when it is satisfied, no matter can appear/disappear or even teleport).
Note that we did not need this condition with the ``particles'' point of view (previous paragraph): of course mass is
conserved if we transport a set of particles with trajectories ${\bf \chi}({\bf x}_0,t)$. Here we need the constraint because
the fluid is represented by two independent fields $\rho$ and ${\bf u}$.

Now let us see how Newton's law (II) can be expressed in terms of $\rho$ and ${\bf u}$. We need to compute acceleration. A first
idea would be to say that acceleration is simply the derivative of ${\bf u}$ with respect to time, but it is wrong ! The acceleration
${\bf a}({\bf x}, t)$ that we want to compute is the acceleration of the particle that passes under grid-point ${\bf x}$ at time $t$. When we move from
$t$ to $t+\delta t$, the particle under the grid-point ${\bf x}$ is replaced by another particle, then a correction needs to be
applied (tracking back in time our initial particle to measure its new speed). The correct formula can be found by computing the
acceleration from the ``particle point of view'' (using the derivatives of ${\bf \chi}$), and mapping them to the correct grid-point:
$$
  {\bf a}({\bf \chi}({\bf x}_0, t),t) = \frac{d^2 {\bf \chi}({\bf x}_0, t)}{d t^2}
$$
  Applying the chain rule, solving the equation and substituting ${\bf x} = {\bf \chi}({\bf x}_0, t)$
(see, e.g., \cite{EmmanuelMaitreCours} for the detailed derivation), one finds:
$$
  {\bf a}({\bf x}, t) = \frac{\partial {\bf u}}{\partial t} + \left[\begin{array}{lll}  {\bf u}_x \frac{\partial {\bf u}_x}{\partial x} \\[2mm]
                                                                                        {\bf u}_y \frac{\partial {\bf u}_y}{\partial y} \\[2mm]
                                                                                        {\bf u}_z \frac{\partial {\bf u}_z}{\partial z} \end{array}\right]
                      = \frac{\partial {\bf u}}{\partial t} + ({\bf u}\cdot\nabla) {\bf u}
$$

This formula, that computes the acceleration of the ``particle under the grid-point'',
is referred to as the ``particle derivative'' or ``material derivative''. We are now equipped to write the equation of the fluid:

$$
\left\{
\begin{array}{lcl}
   \frac{\partial {\bf u}}{\partial t} + ({\bf u}\cdot\nabla) {\bf u} & = & \frac{1}{\rho} ({\bf F} + {\bf P})\\[3mm]
   \frac{\partial \rho}{\partial t} + \nabla \cdot({\rho {\bf u}}) & = & 0
\end{array}
\right.
$$ where:
\begin{itemize}
\item the first equation corresponds to Newton's law (II), written
using the ``particle derivative'' to express acceleration in function
of the velocity field ${\bf u}$. In the right-hand side, {\bf F} denotes
the force (for instance gravity) and the additional force {\bf P} denotes
the pressure (more on this below),
\item the second equation
  corresponds to the conservation of mass (``continuity equation'').
\end{itemize}

To go further, we need now to explicit the pressure force ${\bf P}$, knowing that
is it supposed to enforce non-compressibility. To make things easier, we suppose
that at time $t=0$, the density $\rho$ is uniform (everywhere the same). Then, since
the fluid is incompressible, it remains the same for any time, and we got
$\partial \rho / \partial t = 0$. The continuity equation becomes then
$\nabla \cdot({\rho {\bf u}}) = 0$, or $(\nabla \rho){\bf u} + \rho \nabla \cdot {\bf u} = 0$.
Note that $\rho$ is uniform, thus it has zero gradient everywhere, and the first term vanishes.
Then since $\rho$ is non-zero, we get $\nabla \cdot {\bf u} = 0$. In other words, non-compressibility
implies that ${\bf u}$ has zero divergence. Before going further, I need to tell you more about
this pressure force: it is of the kind that is (minus) the gradient of a potential (pressure force
${\bf P}$ is minus the gradient of the ``pressure field'' $p$). Thus our fluid equation becomes:
$$
\left\{
\begin{array}{lcl}
  \frac{\partial {\bf u}}{\partial t} + ({\bf u}\cdot\nabla) {\bf u} & = & ({\bf F} - \nabla p)\\[3mm]
  \nabla \cdot {\bf u} & = & 0
\end{array}
\right..
$$
In the right hand side of the first equation, I removed the term $1/\rho$ since $\rho$ is uniform and constant
(w.l.o.g. consider it to be 1). Now we can learn more about pressure, by taking the divergence of both sides
of the first equation (Newton's law II). We then obtain:
$$
  \nabla \cdot \frac{\partial {\bf u}}{\partial t} + \nabla \cdot ({\bf u}\cdot\nabla) {\bf u} = \nabla \cdot ({\bf F} - \nabla p)
$$
  Since ${\bf u}$ has zero divergence, the first term ($\nabla \cdot \frac{\partial {\bf u}}{\partial t}$ vanishes. By reordering
  and grouping, and using $\nabla \cdot \nabla p = \Delta p$ (the divergence of the gradient is the Laplacian), one obtains:
$$
  \Delta p = \nabla \cdot ({\bf F} - ({\bf u}.\nabla){\bf u})
$$
Our fluid equation then becomes:
\begin{equation}
\left\{
\begin{array}{lcl}
  \frac{\partial {\bf u}}{\partial t} + ({\bf u}\cdot\nabla) {\bf u} & = & ({\bf F} - \nabla p)\\[3mm]
  \frac{\partial \rho}{\partial t} + \nabla \cdot (\rho {\bf u}) & = & 0 \\[2mm]  
   \Delta p & = & \nabla \cdot ({\bf F} - ({\bf u}.\nabla){\bf u})
\end{array}   
\right..
\label{eqn:EulerPoisson}
\end{equation}
With boundary conditions:
$$
 \left\{
 \begin{array}{llcl}   
   \forall {\bf x} \in \Omega,          & \nabla \cdot {\bf u}({\bf x}, 0)           & = & 0\\
   \forall {\bf x} \in \partial \Omega, & {\bf u}({\bf x}, 0) \cdot {\bf n}({\bf x}) & = & 0
 \end{array}
 \right.
$$
 (the initial velocity field ${\bf u}(.,0)$ has zero divergence, and is tangent to the boundary of the
 domain). There is also a limit condition for the initial pressure field, but it is more tricky (I will
not detail it here). With the additional limit condition, incompressibility ($\nabla \cdot {\bf u} = 0$) can be
removed, and is naturally implied by the fluid equations, that couple the
pressure field with the velocity field, and by the limit conditions.

Let us take a closer look at Equation \ref{eqn:EulerPoisson}. The first equation
expresses the acceleration (particle derivative) in function of the gradient of a potential
(here $\nabla p$), and the second one is a Poisson equation, with the Laplacian of the potential
in its left-hand side. This type of equation is known as a ``Euler-Poisson'' system.

\subsection{Similarities between fluids and astrophysics}
\label{sect:astro}

This subsection explains interesting relations between the reasoning used in the previous section
to establish the equation of fluids and problems in astrophysics that consider self-gravitating
matter, that is, a bunch of stars that mutually attract. This subsection may be skipped in a first
reading. \\

  Euler-Poisson systems can be encountered in different settings. In this subsection, we show an example
in astrophysics. Consider a huge number of stars, each of them having a mass $m_i$ and a position in time
${\bf x}_i(t)$ and a speed ${\bf v}_i(t)$. Knowing the initial positions and speeds, how can we deduce the
movement of all the stars ? One of the difficulties is that each star is attracted by all the other
stars. Now, we will make it even more difficult: since we consider a huge number of stars, we would like to ``take a
step backward'' to look at them from a wider perspective, and consider a continuous density field $\rho$
instead of individual stars. Now since we have a continuous density field, it is natural to represent
the speeds of the stars by a continuous velocity field ${\bf u}$, also attached to fixed locations
in space. How can we write the equation that governs the time evolution of $\rho$ and ${\bf u}$ ?
Again, the difficulty is that a specific point in space is attracted by all the other points (in a way that
depends on both the distance to the other point and the density at the other point). 

To make it easier, let us start with a single mass $M$ at the origin. We take the point of view of another
mass $m$ at point ${\bf x}$ attracted by the mass $M$ at the origin. The mass $m$ ``feels'' a force ${\bf F}$:
$$
   {\bf F} = m {\bf G}(r) = - m {\cal G} M \frac{\bf x}{\left| {\bf x} \right|^3}
$$
   where ${\cal G}$ denotes the universal constant of gravitation. It is easy to check that ${\bf G}$ derives
from a potential $\Phi$, given by:

$$
   {\bf G}({\bf x}) = - \nabla \Phi({\bf x}) \quad ; \quad \Phi({\bf x}) = - {\cal G} \frac{M}{\left| {\bf x} \right|}.
$$

Consider now that the mass $m$ at point ${\bf x}$ is attracted by a continuous field of matter, of density $\rho({\bf x})$.
Then the potential at point ${\bf x}$ is obtained by summing the contributions of all the other points ${\bf x}^\prime$ in
the domain $\Omega$:

\begin{equation}
  \Phi({\bf x}) = \int_\Omega - {\cal G} \frac{\rho({\bf x}^\prime)}{\left| {\bf x} - {\bf x}^\prime \right|} d{\bf x}^\prime
  \label{eqn:GlobalPhi}
\end{equation}

Imagine now that you want to compute $\Phi$ numerically. It will have a high cost, because all points of the domain are
coupled with all the other points of the domain. Then the question is ``is it possible to have an equation for $\Phi$
that comes in local form ?''. By local, I mean that the equation should only involve the value of the density at a given
point ${\bf x}$ and their derivatives. By examining Equation \ref{eqn:GlobalPhi}, it is possible to recognize that the
integrand corresponds to something that is well known, called the ``Green function''. The Green function
$K({\bf x}, {\bf x}^\prime)$ is a way of expressing the solution of a Poisson equation ($\Delta f = g$) as a
convolution, as follows:

$$
\begin{array}{lcl}
  \Delta f({\bf x}) & = & g({\bf x}) \\[2mm]
  f({\bf x}) & = & \int_\Omega K({\bf x}, {\bf x}^\prime) g({\bf x}^\prime) d{\bf x}^\prime \\[2mm]
\end{array}  
$$

The expression of $K$ can be found by solving for $K$ in
$\Delta K({\bf x}, {\bf x}^\prime) = \delta({\bf x} - {\bf x}^\prime)$ where $\delta$ is the Dirac distribution. I'm not
giving the details of the derivations, one may refer to the standard textbooks\footnote{\url{https://en.wikipedia.org/wiki/Green's_function}}. The
expression of $K$ is:

$$
  K({\bf x}, {\bf x}^\prime) = - \frac{1}{4\pi} \frac{1}{\left| {\bf x} - {\bf x}^\prime \right|}.
$$

Comparing with Equation \ref{eqn:GlobalPhi}, one sees that the integrand corresponds to $K$ (up to a constant factor $4\pi{\cal G}$).  Summarizing
what we know so far, for a continuous field of matter $\rho$, the force ${\bf F}(\bf r)$ ``felt'' by a mass $m$ at a point ${\bf x}$ is given by:

$$
\left\{
\begin{array}{lcl}
  {\bf F}({\bf x}) & = & m {\bf G}({\bf x}) = -m \nabla \Phi({\bf x}) \\
  \Delta \Phi & = & 4 \pi {\cal G} \rho
\end{array}
\right..
$$

Now we are ready to write the equations of motion for matter described by a density field $\rho({\bf x}, t)$, and the associated velocity
field ${\bf u}({\bf x},t)$:

\begin{equation}
\left\{
\begin{array}{lcl}
  \frac{\partial {\bf u}}{\partial t} + ({\bf u}\cdot\nabla){\bf u} & = & -\nabla \Phi \\[2mm]
  \frac{\partial \rho}{\partial t} + \nabla \cdot (\rho {\bf u}) & = & 0 \\[2mm]
  \Delta \Phi & = & 4 \pi {\cal G} \rho
\end{array}
\right.
\label{eqn:GravityEulerPoisson}
\end{equation}
where the first equation corresponds to Newton's law II, with in its left-hand side acceleration expressed as the material derivative
and in its right-hand side the gravitational forces as the gradient of the gravitational potential. The second equation is mass
conservation (the continuity equation). The third equation is the Poisson equation (with a Laplacian $\Delta$) that we have just
explained in the previous paragraph. Note that unlike Equation \ref{eqn:GlobalPhi} that is \emph{global}, it only relates $\Phi$ and $\rho$
through \emph{local} relations. As Equation \ref{eqn:EulerPoisson} that we have written for fluids, it is an Euler-Poisson system. We
repeat Equation \ref{eqn:EulerPoisson} below for making it easier to compare both equations:
$$
\left\{
\begin{array}{lcl}
  \frac{\partial {\bf u}}{\partial t} + ({\bf u}\cdot\nabla) {\bf u} & = & ({\bf F} - \nabla p)\\[3mm]
  \frac{\partial \rho}{\partial t} + \nabla \cdot (\rho {\bf u}) & = & 0 \\[2mm]  
  \Delta p & = & \nabla \cdot ({\bf F} - ({\bf u}.\nabla){\bf u})
\end{array}   
\right..
$$

In both equations, the first line corresponds to Newton's law II, with forces that derive from a potential (pressure for fluids,
gravitational potential for astrophysics). The second line, conservation of mass, is the same in both cases. The third line is
a Poisson equation for the potential. The right-hand side is different, in the case of fluids, it connects the pressure with
the velocity fluid, and in the case of astrophysics, it connects the gravitational potential with the density.

In astrophysics, you can simulate the evolution of the whole universe
from big-bang time to now using some form of Equation \ref{eqn:GravityEulerPoisson}, as
done in large-scale simulations such as the DEUS (Dark Energy Universe Simulation)
project\footnote{\url{http://www.deus-consortium.org/}}. The governing equation is similar
to Equation \ref{eqn:GravityEulerPoisson} with some adjustments, (very) roughly summarize here.
See the reference books \cite{Peebles,Hawking}, or the appendix of the article \cite{EUR},
for the detailed derivations. First, it is interesting to
change the coordinates, and make them relative to a global expansion
scale factor, in order to cancel the global effect of expansion and
concentrate on the local aspect. To establish the equations that govern this expansion factor, you
need to take into account the effects of relativity.
Second, you need also to replace the density $\rho$ with its deviation ratio $\rho^\prime$ w.r.t. the average density.
Third, by replacing time $t$ with a function $\tau$ of the global expansion scale factor, you will end up
with a system of equation that is still similar to an Euler-Poisson system for fluids, where the new velocity field
is denoted by ${\bf u}^\prime$ and the new potential by $\varphi$:
$$
\left\{
\begin{array}{lcl}
  \frac{\partial {\bf u}^\prime}{\partial \tau} + ({\bf u}^\prime\cdot\nabla) {\bf u}^\prime & = & - \frac{3}{2 \tau} ({\bf u}^\prime + \nabla \varphi) \\[3mm]
  \frac{\partial \rho^\prime}{\partial \tau} + \nabla \cdot (\rho^\prime {\bf u}^\prime) & = & 0 \\[2mm]
  \Delta \varphi & = & \frac{\rho^\prime - 1}{\tau}
\end{array}
\right..
$$
As you can see, the equation still looks like the original
Euler-Poisson system, with some differences here and there. A
noticeable difference is the right-hand side of the Poisson equation
(third line), with its division by $\tau$. To avoid dividing by zero
at big-bang time $\tau = 0$, the numerator needs to vanish.  This
means, with that model, that density was uniform at big-bang time. Now
the same thing applies to the right-hand side of the first equation,
meaning that at $\tau = 0$, we have ${\bf u}^\prime = -\nabla
\varphi$. In other words, with this model, if you know the potential
$\varphi$ at time $\tau = 0$, then you also know the initial velocity
field ${\bf u}^\prime$ ! (one says that velocity is ``slaved'' to the
potential). This observation (plus some other considerations not
detailed here, refer to \cite{EUR}), is used in methods that attempt
to ``go back in time'' and reconstruct all the trajectories of the
stars from the sole observation of their current locations. Some
examples will be shown in the results section.

\section{Fluids and computers}
\label{sec:FluidNum}

\subsection{Particles or fields ?}

Now that we have seen how to write the equations that describe the motion of a fluid, we will see
how to implement a computer program that implement them. We remind that these equations can be
written in two forms, either you adopt the point of view of a particle (Lagrange coordinates) and write Newton's law (II),
then you get:

\begin{equation}
  \frac{\partial^2 {\bf x}}{\partial t^2} = {\bf F} - \nabla p.
  \label{eqn:FluidParticle}
\end{equation}

Or you adopt the point of view of a person looking at the fluid from a bridge (Euler coordinates), then you obtain:
\begin{equation}
\left\{
\begin{array}{lcl}
  \frac{\partial {\bf u}}{\partial t} + ({\bf u}\cdot\nabla) {\bf u} & = & ({\bf F} - \nabla p)\\[3mm]
  \frac{\partial \rho}{\partial t} + \nabla \cdot (\rho {\bf u}) & = & 0 
\end{array}   
\right.,
\label{eqn:FluidField}
\end{equation}
and with this form you can find an equation that connects the pressure $p$ with the velocity field ${\bf u}$:
$$
  \Delta p = \nabla \cdot ({\bf F} - ({\bf u}.\nabla){\bf u}).  
$$

  Before starting to program anything, you need to make a choice for your variables, either particles or fields. It is
 a dilemma, because each has its pros and cons:
\begin{itemize}
\item With particles, it is interesting because it naturally indicates where the fluid goes, how it deforms etc\ldots
  Adopting this point of view will make it easier to track the geometry of the fluid, and particularly its interfaces
  (free surface, surface between multiple inmiscible fluids). Moreover, it is easier to enforce the conservation of
  the energy with this representation. However, it does not indicate how to compute pressure. The Poisson equation
  that we have for the pressure uses spatial derivatives w.r.t. fixed coordinates. Of course it would be possible
  to translate them to moving particles, but it is non trivial;
\item With fields, pros and cons are the opposite. On the positive side what you get is an easy and natural way of
  computing the pressure. The Poisson equation that gives you the pressure in function of the velocity fluid is
  very classical in numerical analysis, and can be solved very efficiently, especially if you use a regular grid to
  represent the fields. If you use periodic boundary conditions (like in the PacMan game, the fluids that comes out
  from the left side comes in from the right side), then you can use a Fourier transform to solve the equation even
  faster. It is very well explained in the article \cite{DBLP:conf/siggraph/Stam99a} and book \cite{StamBook} by
  Jos Stam, who developed a very efficient algorithm for applications in Computer Graphics.
  On the negative side, the representation of the fluid, as a density and a velocity field,
  is more subject to numerical dissipation, that is, fine structures of the fluid motions that get ``blurred'' during
  the computations. Moreover, conservation laws (conservation of volume, energy, \ldots) are more difficult to enforce.
\end{itemize}   

Several strategies were proposed, to develop numerical schemes that combine the advantage of both coordinate systems.
For instance, PIC (Particle in Cells)
methods \cite{PIC} advect particles and update a pressure field supported by the fixed grid.
These particles can be replaced with more continuous representations, as in SPH methods \cite{SPH}
(Smoothed Particles Hydrodynamics), that derive pressure by replacing the particles with ``Gaussian
splats'', and computing the effect of overlapping splats at a given location in space. It is also possible
to completely free yourself from the need of making a choice, and consider now that you are running along the bank
of the river to follow the fluid. Then you got your own coordinate system (neither Lagrangian nor Eulerian) in which
you translate the equations of motions of the fluid. From a computer point of view, this means that you deform
a grid and make it roughly follow the fluid, in a way that captures the overall geometry of the flow while avoiding
to create degenerated grid elements that would appear if you were following the flow exactly. This class of methods
is called ALE (Arbitrary Lagrangian Eulerian) \cite{ALE}. 

\begin{figure}
 \centerline{
   \includegraphics[width=\textwidth]{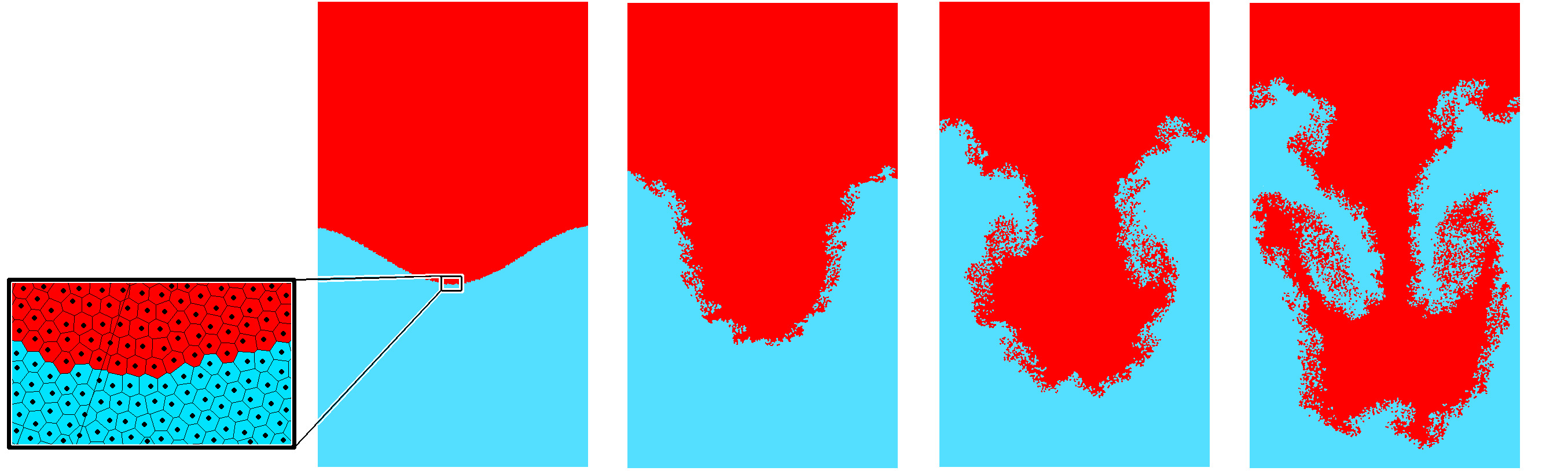}
 }
  \caption{Parameterization of an incompressible fluid using a Laguerre diagram}
\label{fig:laguerre}
\end{figure}

\subsection{A parameterization of incompressible fluids}
\label{sect:W}
Intuitively, the ``particle'' point of view (Lagrangian) in Equation \ref{eqn:FluidParticle} is probably easier to
grasp, because using a computer, you can imagine implementing it with (a variant of) Algorithm \ref{alg:newton1}.
This is the pressure that makes it necessary to introduce a fixed grid (Eulerian) and more complicated computations.
The question is now ``is there a means of computing the pressure field $p$ directly from the position of the particles ?''.

A possibility to answer this question is to define a computer representation of a fluid that depends on a (finite)
set of variables, for instance the $3N$ coordinates of $N$ points ${\bf x}_i$ (black dots in Figure \ref{fig:laguerre}),
and that is defined in such a way that
\emph{any configuration of the variables ${\bf x}_i$ results in an incompressible motion}. This will be the topic of the
rest of this subsection. Once we have it, then the question will be ``how to make it respect Newton's laws ?'', which
will be the topic of the next subsection. \\

To define our computer representation of the fluid, we will decompose the domain $\Omega$ into regions $\Omega_i$ associated
with each point ${\bf x}_i$, that will represent portions of the fluid (the cells displayed in Figure \ref{fig:laguerre}).
A possibility to define such a decomposition is to use the
Voronoi diagram of the $N$ points ${\bf x}_i$, that is, a subdivision of the domain $\Omega$ into regions $\Omega_i$ defined by:
$$
   \Omega_i = \{ {\bf x} \quad | \quad d^2({\bf x}, {\bf x}_i) \le d^2({\bf x}, {\bf x}_j) \quad \forall j ),
$$
where $d^2(.,.)$ denotes the (squared) Euclidean distance.
   
While it is possible to simulate fluid dynamics (and astrophysics) with Voronoi diagrams \cite{DBLP:journals/cse/WhiteS99}, in
our case it will not work, because we want to enforce incompressibility \emph{in the first place}: since each region $\Omega_i$
represents a portion of the fluid, it is supposed to keep the same volume throughout the simulation. With a Voronoi diagram,
there is no reason for the volumes of the regions to remain constant. However, it is possible to use instead a Laguerre diagram
(also called power diagram in the specific case), that depends on an additional vector ${\bf W}$ of $N$ scalars $w_i$. As compared
to a Voronoi diagram, the definition of the cells is slightly modified (uses the $w_i$ coefficients):
$$
   \Omega_i = \{ {\bf x} \quad | \quad d^2({\bf x}, {\bf x}_i) - w_i \le d^2({\bf x}, {\bf x}_j) - w_j \quad \forall j ).
$$

With the additional degrees of freedom $w_i$, it is possible to control the volumes of the cells ! In fact, we are in a very good
situation, because for a given set of ${\bf x}_i$ points, there exists a unique ${\bf W}$ vector (up to a translation) such that
the volumes of the cells match the prescribed values (see \cite{DBLP:conf/compgeom/AurenhammerHA92, gangbo1996, BrenierPFMR91},
or our survey \cite{DBLP:journals/cg/LevyS18} and the references herein).
Moreover, this vector ${\bf W}$ can be easily computed (by maximizing a concave function with a Newton algorithm
\cite{DBLP:journals/corr/KitagawaMT16}). In 3D, the algorithm can be implemented using well-adapted
geometric data structures \cite{journals/M2AN/LevyNAL15}. Now, given the $N$ positions of the points ${\bf x}_i$, you can compute
the vector ${\bf W}$ such that the volume of each cell $\Omega_i$ corresponds to a prescribed value !

This means that we have a computer object, that represents a partition of the fluid into $N$ fluid portions $\Omega_i$. The partition is
parameterized by $N$ points ${\bf x}_i$ of $\Omega$, and for any value of the parameters, the volume of each ``fluid portion''
$\Omega_i$ remains constant. The interesting point with this fluid parameterization is that it has the ``particle point of view'' (Lagrange):
one can track where each individual fluid portion goes, or to track the interface between two inmiscible fluids, as in Figure \ref{fig:laguerre}.
Now we need to see how to use Newton's laws to determine how the points ${\bf x}_i$ should move during the simulation.

\subsection{The Gallouet-Merigot scheme}

In this section, we describe the algorithm for simulating incompressible fluids, due to Gallouet and Merigot \cite{Gallouet2017}. They proved
(after elaborate calculations) that the algorithm converges to the solution of the incompressible Euler equation. Their algorithm can be
also explained from an intuitive point of view:

Let us now consider a fluid subject to gravity. The only force we need to take care of is gravity, there is no-longer pressure since it is
naturally taken into account by our parameterization of the fluid. Thus, the only thing we need to do is applying the effect of gravity
to the ${\bf x}_i$'s. However, some care needs to be taken: remember that each ${\bf x}_i$ has a volume of fluid $\Omega_i$ attached to it.
If we keep that in mind, gravity is applied to \emph{the center of mass of the fluid portion} $\Omega_i$ (and not to ${\bf x}_i$. However, we
cannot directly act on the center of masses of the fluid portions $\Omega_i$, since our only variables are the ${\bf x}_i$'s, but we can attach
a little spring between each ${\bf x}_i$ and the center of gravity of the associated $\Omega_i$. When ``pulling'' ${\bf x}_i$, this will in turn
(and indirectly) pull the center of gravity of $\Omega_i$. Or put differently, if imagining now that our fluid is represented in a purely discrete
manner, as a set of $N$ ${\bf x}_i$ points (that no longer represent fluid portions), then the force exerted by the little spring between ${\bf x}_i$
and the center of gravity of $\Omega_i$ may be thought of as a ``discrete pressure force'', that ensure the incompressibility of the (discrete) fluid.
Putting everything together results in Algorithm \ref{alg:gallouet_merigot}. Compared with our trivial simulation algorithm of the introduction
(Algorithm \ref{alg:newton1}), it just adds the ``pressure force'', that is the ``spring energy'' that connects the points ${\bf x}_i$ to the
centers of gravity of the Laguerre cells ${\bf g}_i$. This simple algorithm simulates interesting behavior of the fluid, such as vortices, as can be seen
in Figure \ref{fig:laguerre} (more results in the next section). One may argue that the function that computes the Laguerre diagrams and the vector ${\bf W}$
(not detailed in this article, refer to \cite{DBLP:journals/cg/LevyS18} and references herein) is far from trivial. However, note that more classical implementations
of fluid simulation, such as \cite{DBLP:conf/siggraph/Stam99a}, also rely on efficient numerical methods (e.g. Fast Fourier Transform). The Laguerre diagram is
less standard than the FFT, but since it solves a form of Optimal-Transport, a problem that is very general, we think that this type of method will make its way
and will be soon part of the standard numerical optimization toolbox.

\begin{algorithm}[t]
\KwIn{Initial positions ${\bf x}_i$ and speeds ${\bf v}_i$ of the $N$ particles}  
  $t \leftarrow 0$\;
  \For{{\rm timestep}=0 {\bf to} {\rm maxstep}} {
    Compute the vector ${\bf W}$ that controls the volume of the Laguerre cells\;
    Compute the Laguerre diagram 
    Compute the centers of gravity ${\bf g}_i$ of the Laguerre cells $\Omega_i$\;
    $t \leftarrow t + \delta t$ \;
    \For{i=1 {\bf to} N} {
       ${\bf F}_i \leftarrow m_i {\bf G} + \frac{1}{\epsilon^2} ({\bf g_i} - {\bf x_i})  $  \tcp*{Update force}
    }
    \For{i=1 {\bf to} N} {    
       ${\bf a}_i \leftarrow \frac{1}{m_i}{\bf F}_i        $  \tcp*{Update acceleration}
       ${\bf v}_i \leftarrow {\bf v}_i + \delta t\ {\bf a}_i$  \tcp*{Update speed}      
       ${\bf x}_i \leftarrow {\bf x}_i + \delta t\ {\bf v}_i$  \tcp*{Update position}
    }
  }
\caption{Simulation of an incompressible fluid with the Gallouet-Merigot algorithm}
\label{alg:gallouet_merigot}
\end{algorithm}

\section{Results}
\label{sec:FluidResults}

\begin{figure}
  \centerline{
    \includegraphics[width=0.33\textwidth]{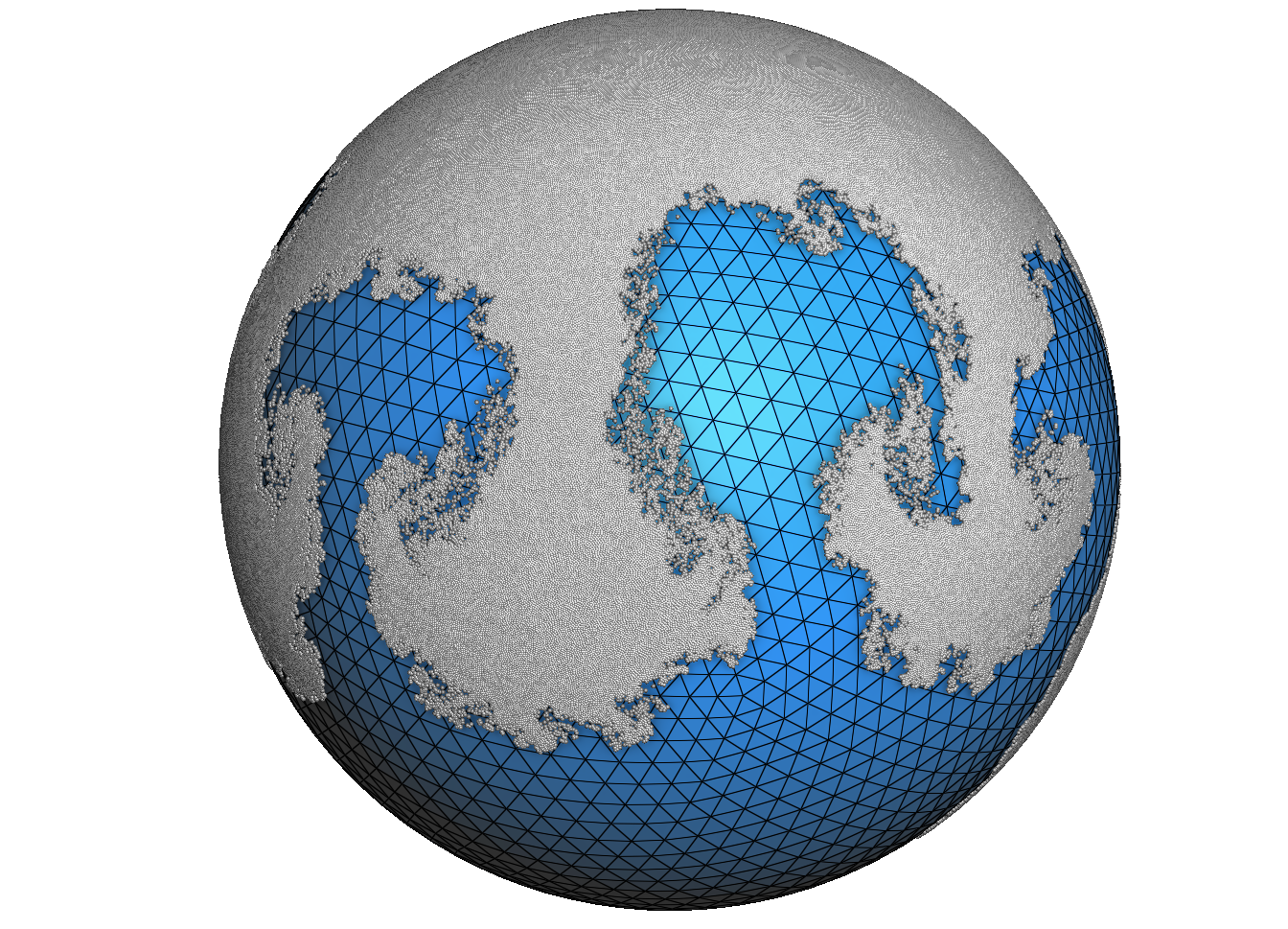}
    \includegraphics[width=0.33\textwidth]{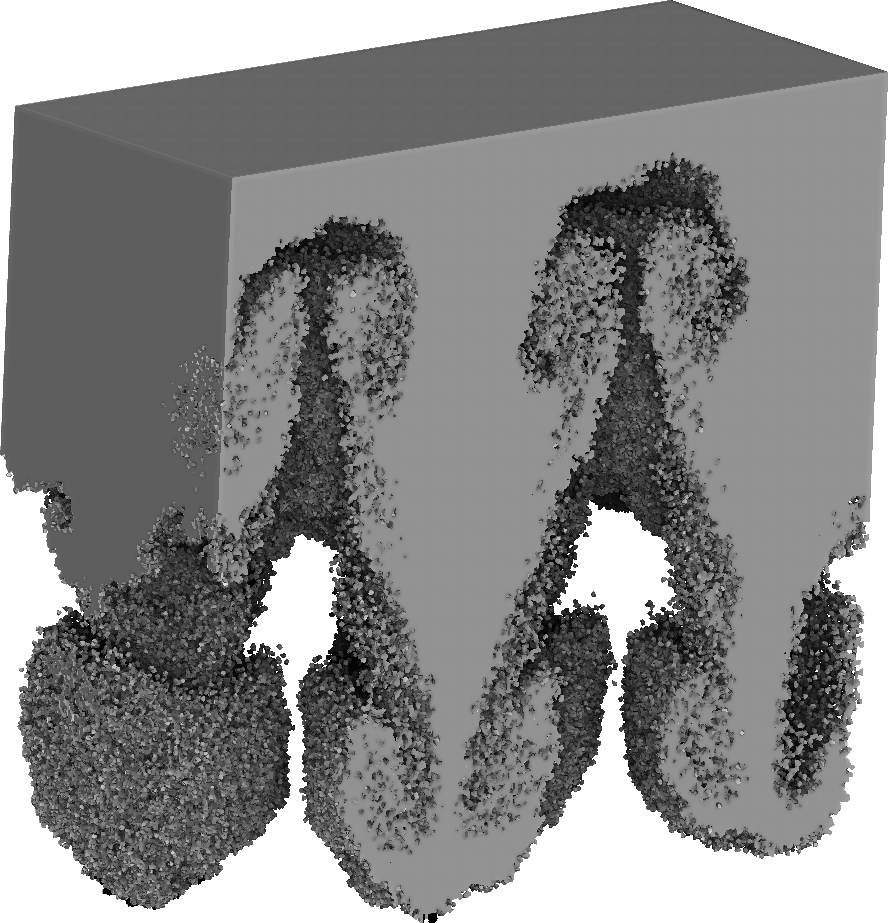}
    \includegraphics[width=0.33\textwidth]{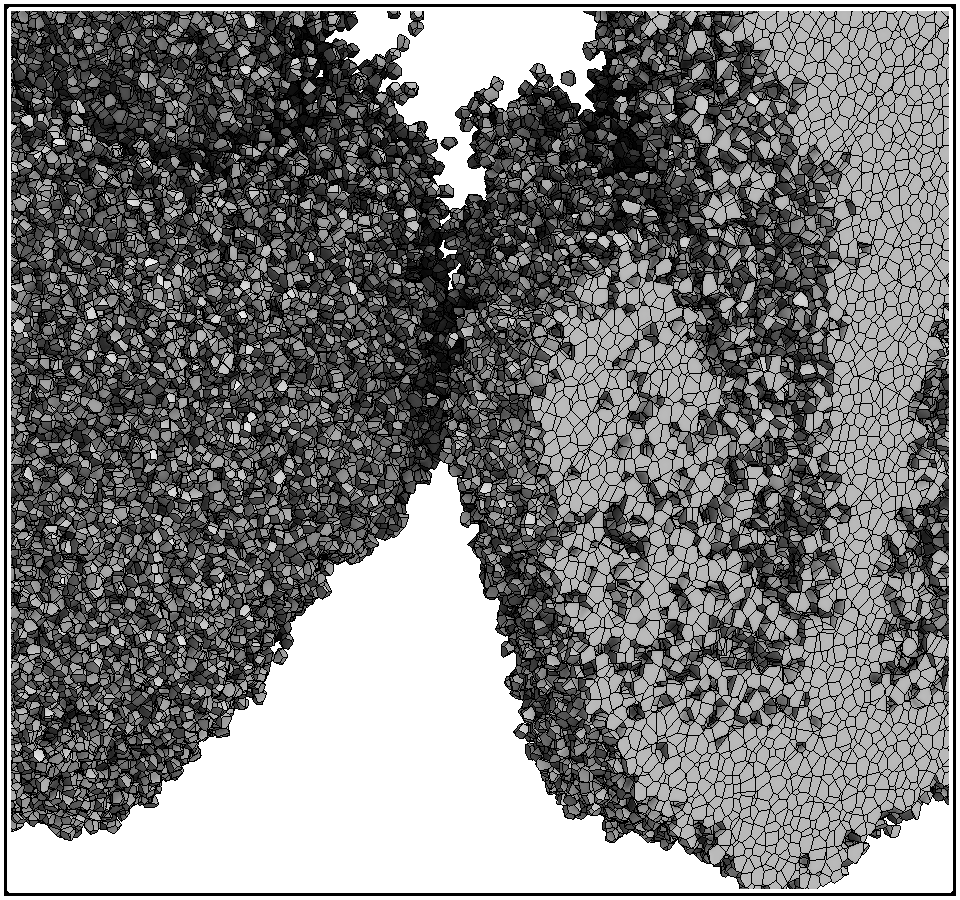}
  }
  \caption{ Fluid on a sphere (left) and 3D instability (right).
  }
\label{fig:TaylorRayleigh3D}
\end{figure}

\begin{figure}
  \centerline{
    \includegraphics[width=\textwidth]{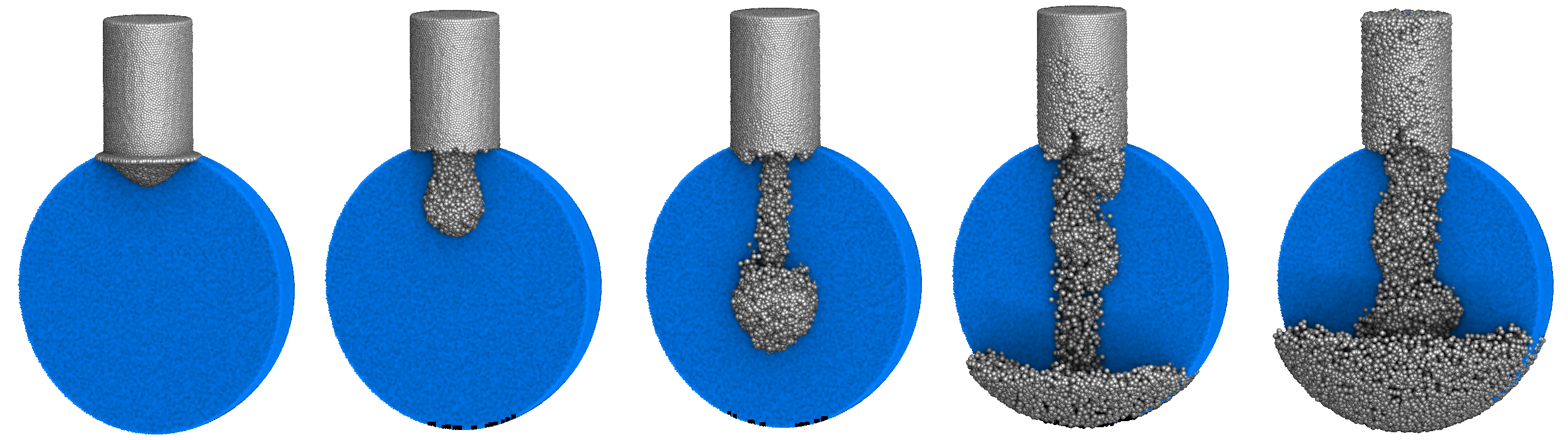}
  }
  \caption{Taylor-Rayleigh instability in a bottle.}
\label{fig:Bottle}
\end{figure}

\begin{figure}
  \centerline{
    \includegraphics[width=\textwidth]{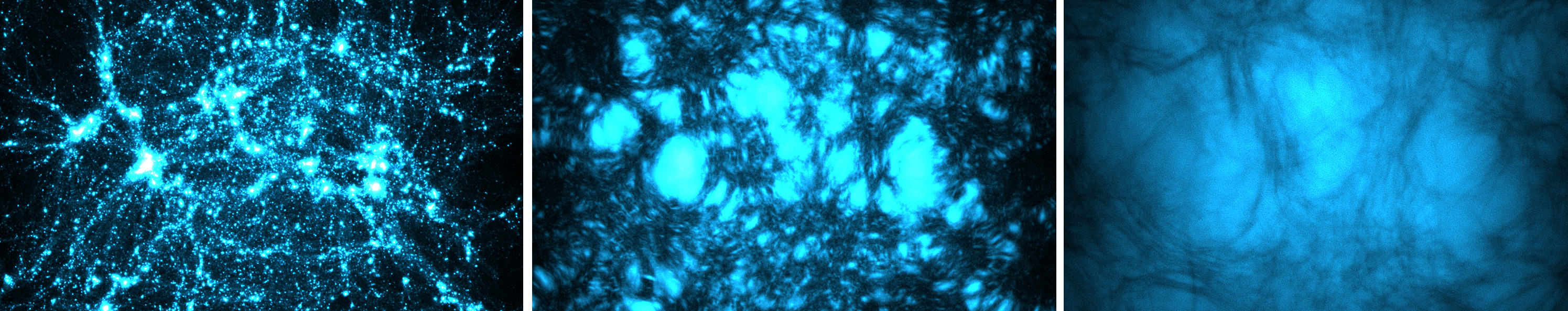}
  }
  \caption{Early Universe Reconstruction (simulation data courtesy of Roya Mohayaee, Institut d'Astrophysique de Paris)}
\label{fig:EUR}
\end{figure}

We show some results. Figure \ref{fig:TaylorRayleigh3D}-left shows a simulation of a fluid constrained
to move on a sphere, that develops vortices. On the right, the same type of vortices, but this time in
full 3D (10 million points). Using efficient geometric algorithms, it is possible to define an arbitrary
shape for the domain $\Omega$, as shown in Figure \ref{fig:Bottle}. Finally, Figure \ref{fig:EUR} shows
some on-going works in Early Universe Reconstruction: starting from the current (simulated for now) positions
of a set of galaxy clusters (left), the goal is to retrieve the initial condition. As explained in Section
\ref{sect:astro}, density is supposed to be constant at the initial time. The question is then to determine
for each galaxy cluster where it took its matter from, by ``inverting'' the equation in Section \ref{sect:astro}.
As our problem of computing the ${\bf W}$ vector that controls the volumes of the Laguerre cells (Section \ref{sect:W}),
this is also an instance of the Optimal Transport problem, that can be solved with the same algorithm (with some adjustments
to make it scalable). Here it is applied to 16 million galaxy clusters, with periodic (``pac-man'') boundary conditions.
Computation took slightly more than 1h on a desktop with 32 Gb RAM and an NVidia V100 for the linear algebra. 

\section*{Conclusions - towards the ``path bundle'' method}

The numerical experiments in this article showed that a Laguerre
diagram with controlled cell volumes is a nice parameterization of an
incompressible fluid, leading to a fluid simulation algorithm that is
straightforward to implement (supposing that you already got the
semi-discrete optimal transport code that computes the Laguerre
diagram, which is not standard, but readily available in the
open-source GEOGRAM
library\footnote{\url{http://alice.loria.fr/software/geogram/doc/html/index.html}}).
Now the way the points ${\bf x}_i$ are connected to the centers of
gravity with a ``spring'' is not completely satisfactory, in
particular it introduces a parameter $\epsilon$ (the stiffness of the
springs).  Could we find instead a way of determining the movement of
the ${\bf x}_i$ in such a way that our computer implementation best
approximates the ``true'' (i.e. continuous, mathematically idealized)
system ? To find the equations of motion of the ${\bf x}_i$'s, one
possibility is considering something else than Newton's laws, called
the principle of least action, that is mathematically equivalent to
Newton's laws. This means starting from the principle of least action,
one retrieves exactly the same equations of motion. But the difference
is that the principle of least action comes in the form of a quantity
(called ``action'', and corresponding to the integrated difference
between the kinetic energy and the potential energy) that is
\emph{minimized}. In our case it makes a difference: we can imagine
parameterizing the action with our discrete variables, and find the
time evolution of our discrete variables that minimize the action. To
do that, it may be possible to use the ``adjoint method'' from optimal
control theory.

\section*{Acknowledgments}
I wish to thank Quentin M\'erigot, Thomas Gallouet,
Yann Brenier, Jean-David Benamou, Roya Mohayaee and Jean-Michel Alimi for many discussions. This research project was
partly funded by the ANR MAGA and INRIA PRE EXPLORAGRAM.

\appendix
\section{The complete sources of a fluid simulator}
This section gives the complete sources of the fluid simulator (shorter than two pages !), implemented in Python. 
The full source code and companion software is available here\footnote{\url{https://gforge.inria.fr/frs/?group_id=1465}}
and the associated video tutorial for installation is here\footnote{\url{https://www.youtube.com/watch?v=2ULkab3vyfc}}. The heart
of the algorithm is the \verb|Euler_step()| function (20 lines !), it is not longer nor not more complicated than Algorithm
\ref{alg:gallouet_merigot}, the rest is initialization and bookkeeping (there are also 30 lines for the GUI that we did not reproduce here because
they are not very relevant for the topic of this article). All 2D images (Taylor-Rayleigh instability) were produced using this code.

\label{apdx:sources}
\begin{verbatim}
import math,numpy
N=1000 # Number of points.

def Euler_step():
   OT = points.I.Transport
   tau = 0.001     # Timestep
   epsilon = 0.004 # Stiffness of the 'spring' pressure force.
   G = 9.81        # Gravity on earth in m/s^2
   inveps2 = 1.0/(epsilon*epsilon)
   # Compute both W and the centroids of the Laguerre diagram.
   OT.compute_optimal_Laguerre_cells_centroids(
      Omega=Omega,centroids=Acentroid,mode='EULER_2D'
   )
   # Update forces, speeds and positions (Explicit Euler scheme, super simple !)
   for v in range(E.nb_vertices):
      # Compute forces: F = spring_force(point, centroid) - m G Z
      Fx = inveps2 * (centroid[v,0] - point[v,0])
      Fy = inveps2 * (centroid[v,1] - point[v,1]) - mass[v] * G
      # V += tau * a ; F = ma ==> V += tau * F / m
      V[v,0] = V[v,0] + tau * Fx / mass[v]
      V[v,1] = V[v,1] + tau * Fy / mass[v]
      # position += tau * V
      point[v,0] = point[v,0] + tau*V[v,0]
      point[v,1] = point[v,1] + tau*V[v,1]
   points.redraw()

def Euler_steps(n):
   for i in range(n):
      Euler_step()

######## Initialization
scene_graph.clear()
Omega = scene_graph.create_object(classname='Mesh',name='Omega')
Omega.I.Shapes.create_square()
Omega.I.Surface.triangulate()
Omega.I.Points.sample_surface(nb_points=N)
scene_graph.current_object = 'points'
points = scene_graph.objects.points
E = points.I.Editor

###### Low level access to point coordinates
point = numpy.asarray(E.find_attribute('vertices.point'))  # [:,[0,1]]

###### Attributes attached to each vertex: 
######    mass, speed vector and centroid of Laguerre cell
mass      = numpy.asarray(E.find_or_create_attribute('vertices.mass'))
V         = numpy.asarray(
             E.find_or_create_attribute(attribute_name='vertices.speed',dimension=2)
            )
Acentroid = gom.create(classname='OGF::NL::Vector',size=E.nb_vertices,dimension=2)
centroid  = numpy.asarray(Acentroid)

##### Initialize masses with nice sine wave, and heavy fluid on top.
for v in range(E.nb_vertices):
   x = point[v,0]
   y = point[v,1]
   f =0.1*math.sin(x*10) 
   if (y-0.5) > f:
      mass[v] = 3
   else:
      mass[v] = 1

# Start with points at centroids, and initial speeds at zero.
def Euler_init():
   OT = points.I.Transport
   OT.compute_optimal_Laguerre_cells_centroids(
     Omega=Omega,centroids=Acentroid,mode='EULER_2D'
   )
   for v in range(E.nb_vertices):
      point[v,0] = centroid[v,0]
      point[v,1] = centroid[v,1]
      V[v,0] = 0.0
      V[v,1] = 0.0 
   points.update()
\end{verbatim}

\bibliographystyle{alpha}
% Set your bibliography file here (without .bib)
\bibliography{fluids}

\end{document}